\documentstyle[aps,amssymb,epsf,twocolumn]{revtex}
\begin{document}
\title{Tomography of Quantum Operations}
\author{G. M. D'Ariano and P. Lo Presti}
\address{Theoretical Quantum Optics Group, INFM Unit\`a di Pavia \\
Dipartimento di Fisica 'Alessandro Volta' -- Universit\`a di Pavia
via Bassi 6, I-27100 Pavia, Italy}
\date{\today}
\maketitle
\begin{abstract}
Quantum operations describe any state change allowed in quantum
mechanics, including the evolution of an open system or the state
change due to a measurement. 
In this letter we present a general method based on
quantum tomography for measuring experimentally the matrix elements of
an arbitrary quantum operation. As input the method needs only a
single entangled state. The feasibility of the technique for the
electromagnetic field is shown, and the experimental setup is
illustrated based on homodyne tomography of a twin-beam.   
\end{abstract}
\date{\today}
\maketitle
\pacs{PACS numbers: }

The typical state change in quantum mechanics is the unitary evolution,
where the final state is related to the initial one via the
transformation $\rho\rightarrow {\cal E}(\rho) \equiv U \rho
U^{\dagger}$, with $U$ unitary operator on the Hilbert space ${\cal
H}$ of the system. Unitary transformations describe only the
evolutions of closed systems, and non-unitary transformations occur
when the quantum system is coupled to an environment or when a
measurement is performed on the system. What is the most general
possible state change in quantum mechanics? The answer is provided by
the formalism of ``quantum operations'' by Kraus \cite{Kraus83a}. Here 
{\em input} and {\em output} states
are connected via the map 
\begin{eqnarray}
\rho \rightarrow 
\frac{{\cal E}(\rho)}{\mbox{Tr}\bigl({\cal E}(\rho)\bigr)}\;.\label{map}
\end{eqnarray}
The {\em quantum operation\/} ${\cal E}$ is a linear, trace-decreasing
map that preserves positivity (more precisely the map must be
completely positive (CP) \cite{note1}). The trace in the denominator
is included in order to preserve the normalization $\mbox{Tr}(\rho)
= 1$. The most general form for ${\cal E}$ can be shown to be
\cite{Kraus83a} 
\begin{eqnarray}
{\cal E}(\rho) = \sum_n K_n \rho K_n^{\dagger}\;,\label{kraus-mix} 
\end{eqnarray}
where the operators $K_n$ satisfy the bound
\begin{eqnarray}
\sum_n K_n^\dagger K_n\le I\;.\label{sum}
\end{eqnarray}
The transformation (\ref{kraus-mix}) occurs with generally non-unit
probability $\mbox{Tr}\bigl({\cal E}(\rho)\bigr)\le 1$, and the
probability is unit only when ${\cal E}$ is trace-preserving, i.e. when 
the bound (\ref{sum}) is achieved with the equal sign. The particular
case of unitary transformations corresponds to having only one term $K_1 = U$
in the sum (\ref{kraus-mix}), with $U$ unitary. However, one can consider also
non-unitary operations with only one term, i. e.   
\begin{eqnarray}
{\cal E}(\rho) = A \rho A^{\dagger}\;,\label{kraus-pure} 
\end{eqnarray}
with $A$ a {\em contraction}, i. e. $||A||\le 1$: we'll call 
these last operations {\em pure}, since they leave pure states $\rho$ as
pure. Indeed, for $\rho=|\varphi\rangle\langle\varphi |$ we can rewrite
Eq. (\ref{map}) in the form 
\begin{eqnarray}
|\varphi\rangle\rightarrow \frac{A|\varphi\rangle}{||A|\varphi\rangle||}
\;\label{psimap}
\end{eqnarray}
Such an operation could, for example, describe the state
reduction from a measurement apparatus for a given fixed outcome,
which occurs with probability $\mbox{Tr}\bigl(\rho A^{\dagger}
A\bigr)\le 1$. 

Suppose now that we have a quantum machine that performs an unknown
quantum operation ${\cal E}$, and we want to determine ${\cal E}$
experimentally. This problem has been posed in several papers, with 
solutions given in some special cases \cite{turchette,poyatos,jones}.

How can we do? This would be the case, for example, if we
want to determine the unitary transformation $U$ performed by a quantum
device, or the state-reduction achieved by a measuring apparatus that
performs an indirect measurement on the system. In Refs.  
\cite{nielsen,macca} as a method it was suggested to carry on a
tomographic reconstruction at the machine output for a varying
input state. However, the availability of all possible input states
is a practically unsolvable problem. For example, the method of
Ref. \cite{macca} in the optical domain works only for
phase-insensitive devices, since for phase-sensitive ones 
one would need input superpositions of two photon-number states,
superpositions which are currently not feasible. As we will show in
this letter, we can exploit the {\em quantum parallelism} of
entanglement\cite{popescubook} to run all possible input states in
parallel using only a single entangled state as the input in the
tomographic reconstruction. In this way we have at our disposal a
general method for experimentally determining the quantum operation matrix,
using any available quantum-tomographic scheme for the system in
consideration, and a single fixed state at the input, which is an
entangled (not even maximally) state. In the optical domain we will 
show that one can achieve the tomographic reconstruction of the operation
using exactly the same apparatus of the recently performed
experiment of Ref.\cite{kumarandme}.  

Let's consider for simplicity a ``pure'' quantum operation in the form 
(\ref{psimap}). Given an orthonormal basis $\{|j\rangle\}$ corresponding
to some physical observable, how can we determine the matrix
$A_{ij}=\langle i|A|j\rangle$ experimentally? Instead of acting with
the contraction $A$ on a ``isolated'' system, we perform the map on a
system which is entangled in the state $|\psi\rangle\!\rangle\in{\cal
H}\otimes{\cal H}$ with an identical system, i. e.
\begin{eqnarray}
|\psi\rangle\!\rangle\rightarrow |\phi\rangle\!\rangle =
\frac{A\otimes I|\psi\rangle\!\rangle}{||A\psi||_{HS}}\;.\label{psi2map}
\end{eqnarray}
With the double ket we denote bipartite vectors
$|\psi\rangle\!\rangle\in{\cal H}\otimes{\cal H}$, which, keeping the 
basis $\{|j\rangle\}$ as fixed, are in one-to-one correspondence with
matrices as follows 
\begin{eqnarray}
|\psi\rangle\!\rangle=\sum_{ij}\psi_{ij}|i\rangle\otimes|j\rangle\;.
\label{matrixform} 
\end{eqnarray}
In the following we will also use the simple notation of using the
same symbol $A$ for both the matrix $A=\{A_{ij}\}$ and the corresponding
operator $A=\sum_{ij}A_{ij}|i\rangle\langle j|$ for fixed basis 
$\{|j\rangle\}$. With this notation the norm $||A||_{HS}$ in
Eq. (\ref{psi2map}) denotes the Hilbert-Schmidt norm 
$||A||_{HS}=\left[\mbox{Tr}\left(A^{\dagger}A\right)\right]^{{1\over2}}$.
We'll also denote by $A^*$ the operator corresponding to the complex
conjugated matrix of $A$ (with respect to the same fixed basis
$\{|j\rangle\}$), and analogously $A^T$ will denote the
transposed-matrix operator. With consistent notation we'll write
$A=\{A_{ij}\}\equiv[A(j)]$ to denote the column vectors 
$A(j)$ of the matrix $A$, and use 
$|A(j)\rangle=A|j\rangle\equiv\sum_iA_{ij}|i\rangle$ for the
corresponding vectors in ${\cal H}$.  
Using this simple formalism, the quantum operation
matrix $A$ in terms of the input and output state-matrices  writes as follows
\begin{eqnarray}
A=\phi\psi^{-1}\sqrt{p_A(\psi)}\;,\label{A1}
\end{eqnarray}
where $p_A(\psi)=||A\psi||^2_{HS}$ denotes the occurrence probability
of the quantum operation, and the entangled state is assumed to have
invertible matrix $\psi$ (which is always the case in practice). In
our matrix formalism the matrix $\phi$ corresponding to the output
state can be written in terms of measurable ensemble averages as
follows  
\begin{eqnarray}
\phi_{ij}\equiv\langle\!\langle i,j|\phi\rangle\!\rangle=
e^{i\theta}\frac{\langle\, |i_0,j_0\rangle\!\rangle
\langle\!\langle i,j|\,\rangle}{\sqrt{\langle\,
|i_0,j_0\rangle\!\rangle\langle\!\langle i_0,j_0|\,\rangle}}\;, 
\end{eqnarray}
where $\langle
\ldots\rangle\equiv\langle\!\langle\phi|\ldots|\phi\rangle\!\rangle$ denotes 
the ensemble at the output,
$|i,j\rangle\!\rangle\equiv |i\rangle\otimes|j\rangle$, $i_0,j_0$ are
suitable fixed integers, and $e^{i\theta}$ is an irrelevant 
(unmeasurable) overall phase factor corresponding to
$\theta=\arg({\langle\!\langle i_0,j_0|\phi\rangle\!\rangle})$.   
Using Eq. (\ref{A1}) we can write the matrix $A_{ij}$ in terms of 
only output ensemble averages as follows
\begin{eqnarray}
A_{ij}=\kappa \langle E_{ij}(\psi)\rangle\;,\label{estimation}
\end{eqnarray}
where the operator $E_{ij}(\psi)$ is given by
\begin{eqnarray}
E_{ij}(\psi)=
|i_0\rangle\langle i|\otimes|j_0\rangle\langle\psi^{-1*}(j)|\;,\label{Eij}
\end{eqnarray}
and the proportionality constant is given by 
\begin{eqnarray}
\kappa=e^{i\theta}\sqrt{\frac{p_A(\psi)}{\langle\,
|i_0,j_0\rangle\!\rangle\langle\!\langle i_0,j_0|\,\rangle}}\;.\label{kappa}
\end{eqnarray}
Since $A_{ij}$ is written only in terms of output ensemble averages, it
can be estimated through quantum tomography. 
Quantum tomography\cite{gentomo_ortho} is a method to estimate the
ensemble average $\langle H\rangle$ of any arbitrary operator $H$ on
${\cal H}$ by using only measurement outcomes of a {\em quorum} of
observables $\{ O(l)\}$.  
A {\em quorum} is just a set of operators $\{ O(l)\}$ which are
observable (i.e. have orthonormal resolution), and span the linear
space of operators on ${\cal H}$. This means that any operator $H$ can 
be expanded as $H=\sum_l\mbox{Tr}[Q^{\dag}(l)H] O(l)$, where $\{Q(l)\}$ and
$\{O(l)\}$ form a biorthogonal set such that
$\mbox{Tr}[Q^{\dag}(i)O(j)]=\delta_{ij}$. Hence, the tomographic
estimation of the ensemble average $\langle H\rangle$ is obtained as
the double average---over both the ensemble and the quorum---of the
unbiased estimator $\mbox{Tr}[Q^{\dag}(l) H]O(l)$ with random $l$. The
most popular example of quantum tomography is homodyne
tomography\cite{bilkent}, where the 
quorum (selfdual) is given by the operators $\exp(ikX_{\phi})$ for
varying $k$ and $\phi$, $X_{\phi}$ denoting a quadrature of one mode 
of radiation. Notice that for estimating the density matrix also the
maximum-likelihood strategy can be used instead of
averaging\cite{maxlik,tomogroup}. Moreover, there is a general
method\cite{tomogroup} for deconvolving instrumental noise when
measuring the quorum, which resorts to finding the biorthogonal basis
for the noisy quorum. This is the case, for example, of deconvolution
of noise from non-unit quantum efficiency in homodyne
tomography\cite{bilkent}. Finally, for multipartite quantum systems,
one can simply use as a quorum the tensor product of single-system
quorums\cite{tomogroup}: this means that, in our case, we just need to
make two local quorum measurements jointly on the two systems, and
analyze data with the tensor-product estimators. For example, the
estimation of $A_{ij}$ in Eq. (\ref{estimation}) resorts to the calculation
of the following ensemble average from the experimental data
\begin{eqnarray}
A_{ij}=\Bigg\langle\kappa\sum_{kl}a_{ij}(kl)O(k)\otimes O(l)\Bigg\rangle\;,
\end{eqnarray}
where the c-numbers $a_{ij}(kl)$ are given by
\begin{eqnarray}
a_{ij}(kl)=\langle i|Q^{\dag}(k)|i_0\rangle\langle
\psi^{-1*}(j)|Q^{\dag}(l)|j_0\rangle\;.
\end{eqnarray}
Also the fixed ensemble average $\langle\,
|i_0,j_0\rangle\!\rangle\langle\!\langle i_0,j_0|\,\rangle$ in the 
constant $\kappa$ can be measured via tomography, or even by
coincidence counting, whereas $p_A(\psi)$ results from counting the
occurrence of $A$ (occurrence is checked by reading the apparatus,
e. g. $A$ is in correspondence with a given measurement outcome). 

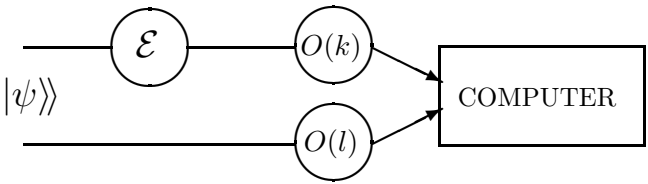
\begin{figure}[hbt]
\vskip .5truecm
\begin{center}
\setlength{\unitlength}{800sp}%
\begingroup\makeatletter\ifx\SetFigFont\undefined%
\gdef\SetFigFont#1#2#3#4#5{%
  \reset@font\fontsize{#1}{#2pt}%
  \fontfamily{#3}\fontseries{#4}\fontshape{#5}%
  \selectfont}%
\fi\endgroup%
\begin{picture}(19506,5446)(1549,-5109)
\thicklines
\put(1801,-3961){\line( 1, 0){8400}}
\put(12601,-961){\vector( 2,-1){2160}}
\put(12601,-3961){\vector( 2, 1){2160}}
\put(1801,-961){\line( 1, 0){2700}}
\put(6901,-961){\line( 1, 0){3300}}
\put(14701,-3961){\framebox(6300,3000){}}
\put(17701,-2761){\makebox(0,0)[b]{\smash{\SetFigFont{10}{49.2}{\rmdefault}{\mddefault}{\updefault}COMPUTER}}}
\put(11326,-1186){\makebox(0,0)[b]{\smash{\SetFigFont{11}{49.2}{\rmdefault}{\mddefault}{\updefault}$O(k)$}}}
\put(11326,-4186){\makebox(0,0)[b]{\smash{\SetFigFont{11}{49.2}{\rmdefault}{\mddefault}{\updefault}$O(l)$}}}
\put(5626,-1261){\makebox(0,0)[b]{\smash{\SetFigFont{15}{49.2}{\rmdefault}{\mddefault}{\updefault}${\cal E}$}}}
\put(2026,-2761){\makebox(0,0)[b]{\smash{\SetFigFont{15}{49.2}{\rmdefault}{\mddefault}{\updefault}$|\psi\rangle\!\rangle$}}}
\put(5701,-961){\oval(2372,2372)}
\put(11401,-886){\oval(2372,2372)}
\put(11401,-3886){\oval(2372,2372)}
\end{picture}
\end{center}
\caption{General experimental scheme of the method for the tomographic
estimation of a quantum operation. Two identical quantum systems are
prepared in an entangled state $|\psi\rangle\!\rangle$. One of the two systems
undergoes the quantum operation ${\cal E}$, whereas the other is left
untouched. At the output one makes a quantum tomographic estimation,
photocurrent by measuring jointly two random observables from a quorum $\{O(l)\}$
(see the text).} \label{scheme}\end{figure}

The general experimental scheme of the method for the tomographic
estimation of a quantum operation matrix is sketched in
Fig. \ref{scheme}. 

The method given above can be easily generalized to the case of
arbitrary non pure quantum operation, as in Eqs. (\ref{map}) and 
(\ref{kraus-mix}). Now the output state is the joint density matrix
\begin{eqnarray}
|\psi\rangle\!\rangle\langle\!\langle\psi|&\rightarrow&
R(\psi)\equiv {\cal E}\otimes{\cal
I}(|\psi\rangle\!\rangle\langle\!\langle\psi|) \nonumber\\
&\equiv&\sum_n K_n\otimes I |\psi\rangle\!\rangle\langle\!\langle\psi|
K^{\dagger}_n\otimes I\;.
\end{eqnarray}
One can immediately see that the quantum operation can be written in
terms of the density matrix $R(\psi)$ for $\psi=I$, i. e.
\begin{eqnarray}
{\cal E}(\rho)=\mbox{Tr}_2[I\otimes\rho^T R(I)]\;,\label{CP}
\end{eqnarray}
where $\mbox{Tr}_2$ denotes the partial trace on the second Hilbert
space. However, for invertible $\psi$ the two matrices $R(I)$ and
$R(\psi)$ are connected as follows
\begin{eqnarray}
R(I)=(I\otimes\psi^{-1T})R(\psi)(I\otimes\psi^{-1*})\;.
\end{eqnarray}
Hence, the (four-index) matrix $R$ in Eq. (\ref{CP}) which is in
one-to-one correspondence with the quantum operation ${\cal E}$ can be 
obtained by estimating via quantum tomography the following output
ensemble averages 
\begin{eqnarray}
\langle\!\langle i,j|R(I)|l,k\rangle\!\rangle
&=&\big\langle\; E^{\dag}_{lk}(\psi)E_{ij}(\psi)\;\big\rangle=\nonumber\\
&=&\big\langle\;|l\rangle\langle i|\otimes
|\psi^{-1*}(k)\rangle\langle \psi^{-1*}(j)|\;\big\rangle\;.
\label{estimationCP}
\end{eqnarray}

We now analyze the experimental feasibility of the method in the
optical domain, based on tomographic homodyning a twin-beam from
parametric downconversion of the vacuum. As a simple example of
quantum operation we consider the unitary displacement
$D(z)=\exp(za^{\dag}-z^*a)$, of a single radiation mode with
annihilation and creation operators $a$ and $a^{\dag}$. The
experimental apparatus is the same as in the experiment of
Ref. \cite{kumarandme}, with a nondegenerate 
optical parametric amplifier (a KTP crystal) pumped by the second
harmonic of a Q-switched mode-locked Nd:YAG laser, which produces a
100-MHz train of 120-ps duration pulses at 1064 nm. The orthogonally
polarized twin beams emitted by the KTP crystal (one of which is
displaced of $D(z)$ by a nearly transparent beam splitter with a
strong local oscillator) are separately detected by two balanced
homodyne setups that use two independent local oscillators derived
from the same laser, with the amplified output noise  at 
radio-frequencies downconverted to the near-dc by use of an rf mixer
and sampled by a boxcar integrator. The outputs of the boxcar channels
are a measure of the quadrature amplitudes $X_{\phi'}\otimes
X_{\phi''}$ for random phases $\phi'$ and $\phi''$  with respect to
the local oscillators, where the quadratures
$X_{\phi}={1\over2}(a^{\dag}e^{i\phi} +ae^{-i\phi})$ here represent
the quorum of observables for the tomographic reconstruction (for
additional details on the experimental setup see
Ref. \cite{kumarandme}, whereas for a more extensive theoretical
treatment see Ref. \cite{twoself}). 

\begin{figure}[hbt]\begin{center}
\epsfxsize=.5\hsize\leavevmode\epsffile{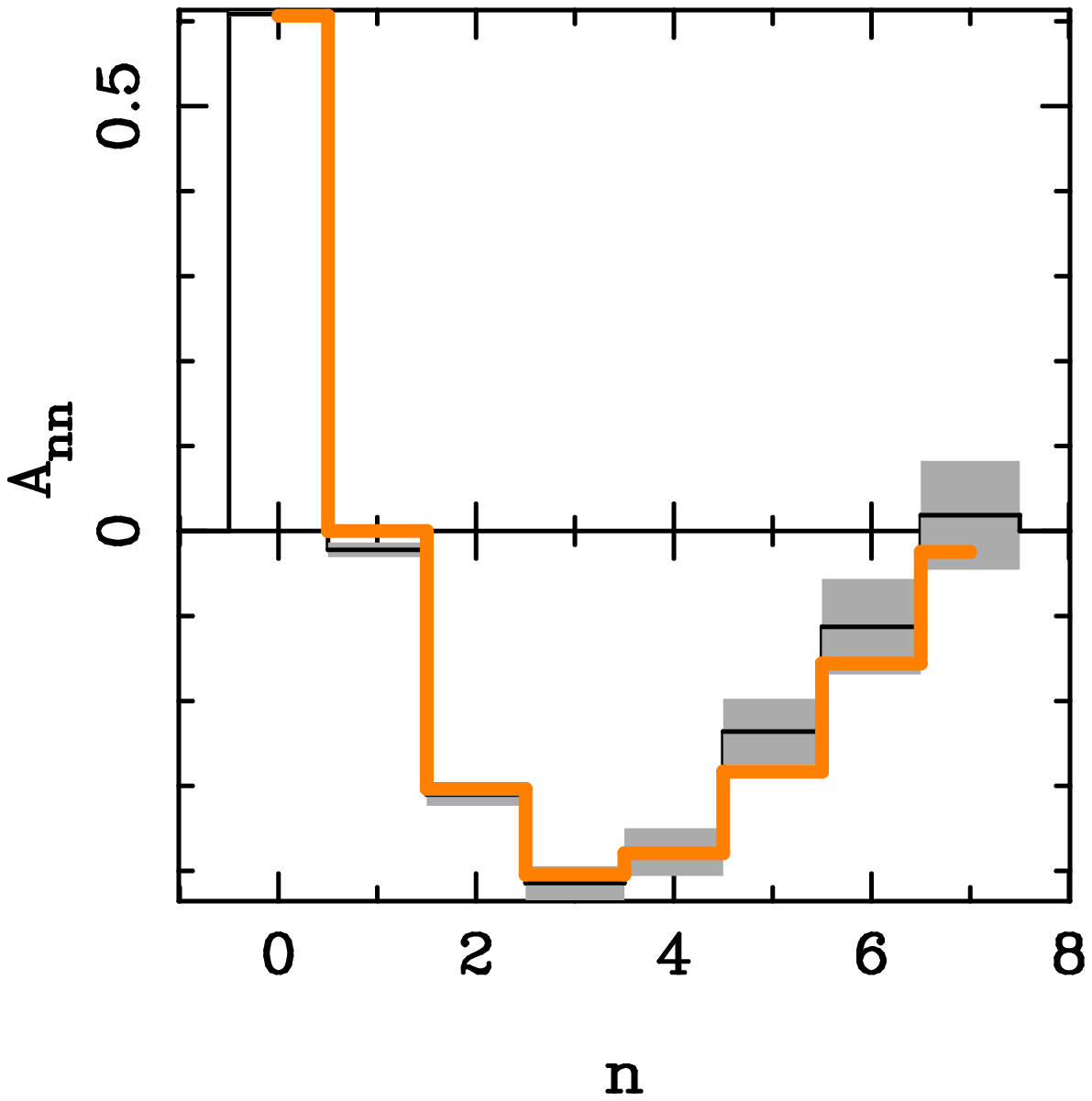}
\epsfxsize=.5\hsize\leavevmode\epsffile{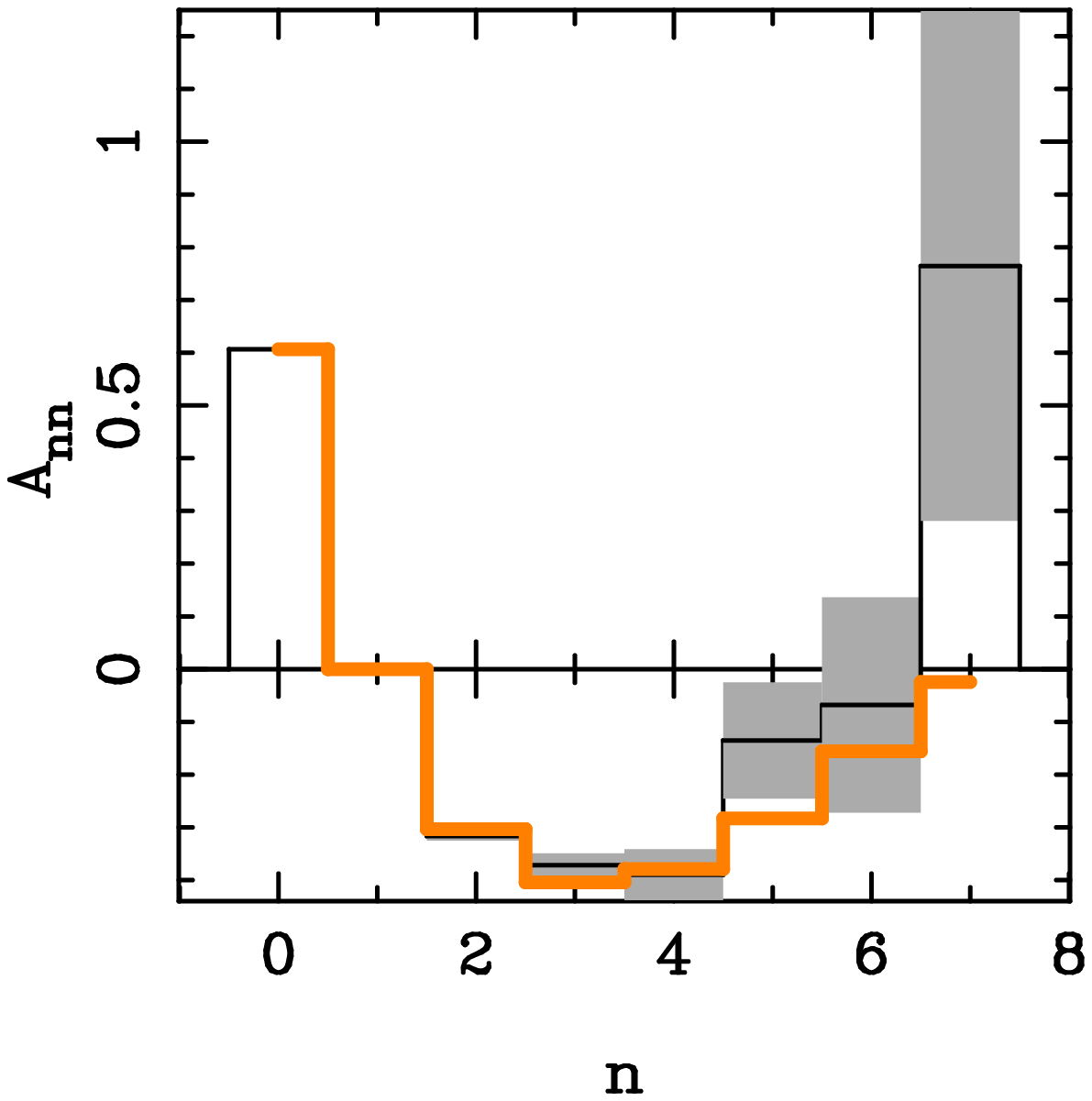}
\end{center}
\caption{Homodyne tomography of the quantum operation $A$
corresponding to the unitary displacement of one mode of the
radiation field. Diagonal elements $A_{nn}$ (shown by thin solid line
on an extended abscissa range,) with their respective error bars in
gray shade, compared to the theoretical probability (thick solid
line). Similar results are obtained for all upper and lower diagonals
of the quantum operation matrix $A$. The reconstruction has been
achieved using an entangled state $|\psi\rangle\!\rangle$ at the input
corresponding to parametric downconversion of vacuum with mean thermal
photon $\bar n$ and quantum efficiency at homodyne detectors
$\eta$. Top: $z=1$, $\bar n=5$, $\eta=0.9$, and $150$ blocks of
$10^4$ data have been used. Bottom: $z=1$, $\bar n=3$,
$\eta=0.7$, and $300$ blocks of $2\cdot 10^5$ data have been used.
The bottom plot corresponds to the same parameters of the experiment
in Ref.\protect\cite{kumarandme}.}\label{simul}\end{figure}  

In Fig. \ref{simul} the results from a homodyne tomography of an
optical displacement of one of the two twin beams from parametric
downconversion of the vacuum are presented for a simulated experiment, 
for displacement parameter $z=1$, and for some typical
values of the quantum efficiency $\eta$ at homodyne detectors
and of the mean thermal photon number $\bar n$ of the twin beam.
As one can
see a meaningful reconstruction of the matrix can be achieved in the
given range with $10^6\div10^7$ data, but this number can be decreased
of a factor $100-1000$ using the tomographic max-likelihood techniques of
Ref. \cite{maxlik}, however at the expense of the complexity of the
algorithm. Homodyne overall quantum efficiencies and amplifier gains
(for the twin-beam) typical of the experimental setup of
Ref. \cite{kumarandme} are considered. Improving quantum
efficiency and increasing the amplifier gain (toward a maximally 
entangled state) have the effect of making statistical errors smaller and
more uniform versus the photon labels $n$ and $m$ of the matrix
$A_{nm}$. Meaningful reconstructions can be achieved with as few as
$\bar n\sim 1$ thermal photons, and with quantum efficiency as low as
$\eta=0.7$. 

We want to mention that the present quantum tomographic method for
measuring the matrix of a quantum operation can be much improved 
by means of a max-likelihood strategy aimed at the estimation of some
unknown parameters of the quantum operation (such max-likelihood
strategy should not be confused with the max-likelihood method of
for the tomographic reconstruction in Ref. \cite{maxlik}). In this case,
instead of obtaining the matrix elements of $R(I)$ from the 
ensemble averages in (\ref{estimationCP}), one has $R(I)$ parametrized
in terms of unknown quantities to be experimentally determined, and
the likelihood is maximized for the set of experimental
data at various randomly selected (tensor) quorum elements, keeping
the same fixed entangled input state. This method is especially
useful for a very precise experimental comparisons between the
characteristics of a given device (e.g. the gain and loss of an active
fiber) with those of a {\em quantum standard} reference \cite{our}.  

In conclusion, in this letter we have presented a general tomographic
method for measuring the matrix of any quantum operation of arbitrary
quantum system. The method exploits the quantum parallelism of
entanglement, with a single entangled state playing the role of a
varying input state, thus overcoming the practically unsolvable
problem of availability of all possible input states for the
tomographic analysis of the quantum operation. We have shown the
feasibility of the method for the case of the electromagnetic field
via homodyne tomography of a twin-beam from nondegenerate
downconversion of the vacuum. The unilateral displacement of the
twin-beam has been considered, and for displacement parameters of the
order of unit our results show that the tomographic estimation can be
achieved using the same apparatus of a similar recently performed
experiment. 

\par This work has been supported by the Italian Ministero
dell'Universit\`a e della Ricerca Scientifica e Tecnologica (MURST)
under the co-sponsored project 1999 {\em Quantum Information
Transmission And Processing: Quantum Teleportation And Error
Correction}. 


\begin{thebibliography}{10}
\bibitem{Kraus83a} K. Kraus, {\em States, Effects, and Operations}
(Springer-Verlag, Berlin, 1983).
\bibitem{note1} A map ${\cal E}$ is completely positive if it preserves
positivity not just for a local state in ${\cal H}$, but also for any
state of the system that is entangled with any other system. In other
words, upon denoting by ${\cal I}$ the identical map on the Hilbert
space ${\cal K}$ of a second quantum system, the extended map ${\cal
E}\otimes{\cal I}$ on ${\cal H}\otimes{\cal K}$ must be positive for
any extension ${\cal K}$. 
\bibitem{turchette} Q. A. Turchette, C. J. Hood, W. Lange, H. Mabuchi,
and H. J. Kimble, Phys. Rev. Lett. {\bf 75} 4710 (1995)
\bibitem{poyatos} J. F. Poyatos, J. I. Cirac, and P. Zoller
Phys. Rev. Lett. {\bf 78} 390 (1997) 
\bibitem{jones} K. R. W. Jones, Phys. Rev. A {\bf 50} 3682 (1994) 
\bibitem{nielsen} I. L. Chuang and M. A. Nielsen, J. Mod. Opt. {\bf
44} 2455 (1997) 
\bibitem{macca} G. M. D'Ariano and L. Maccone, Phys. Rev. Lett. {\bf
80} 5465 (1998)  
\bibitem{popescubook} {\em Introduction to Quantum Computation and
Information}, Ed. by H.-K. Lo, S. Popescu, T. Spiller, World
Scientific (Singapore 1998).  
\bibitem{kumarandme} M. Vasilyev, S.-K. Choi, P. Kumar, and G. M. D'Ariano,
Phys. Rev. Lett. {\bf 84} 2354 (2000)
\bibitem{gentomo_ortho} For a recent paper on quantum tomography
presenting the general approach of the present paper, along with
examples, see: G. M. D'Ariano, L. Maccone, and M. G. A. Paris,
Phys. Lett. A {\bf 276} 25 (2000). For a starting point on the
extensive literature on this topic see references therein. 
\bibitem{bilkent} For a review, see: G. M. D'Ariano, {\em Measuring
Quantum States}, in {\em Quantum Optics and Spectroscopy of Solids},
ed. by T. Hakio\v{g}lu and A. S. Shumovsky, (Kluwer Academic
Publisher, Amsterdam 1997), p. 175-202
\bibitem{maxlik} K. Banaszek, G. M. D'Ariano, M. G. A. Paris,
M. Sacchi, {\em Maximum-likelihood estimation of the density matrix},
Phys. Rev A {\bf 61}, 010304 (2000) (rapid communication)
\bibitem{tomogroup} G. M. D'Ariano, {\em Universal quantum 
estimation}, Phys. Lett. A {\bf 268} 151 (2000)
\bibitem{twoself} G. M. D'Ariano, M. Vasilyev, and Prem Kumar {\em
Self-homodyne tomography}, Phys. Rev. A {\bf 58} 636 (1998)
\bibitem{our} G. M. D'Ariano, P. Lo Presti, and M. G. A. Paris, unpublished.
\end{thebibliography}
\end{document}